\documentclass{article}
\begin{document}
\title{ 
Light Quark Spectroscopy and Charm Decays
}
\author{
 Alberto Reis\\ 
Centro Brasileiro de Pesquisas F\'{\i}sicas\\
\em Rua Dr. Xavier Sigaud, 150 - Rio de Janeiro - Brazil
}
\maketitle
\baselineskip=11.6pt
\begin{abstract}
The connection between light quark spectroscopy and hadronic decays of D mesons
is discussed, with emphasis on the physics of the light scalar mesons. Recent 
results from charm decays are presented.   
\end{abstract}
\baselineskip=14pt
\section{Introduction}

Forty years have passed since the birth of the Constituent Quark Model (CQM).
This model provided a very successful description of almost all the hadronic 
spectrum. The nonets of pseudo-scalar, vector and tensor mesons are now well
identified. There is, however, one remaining and crucial problem: 
the identification of the scalar meson nonet(s). The solution of this enigma 
is of vital importance for understanding QCD at the low energy limit.

On the other side, there has been tremendous progress in charm physics
in the past decade. High quality data allowed the basic properties of charm 
mesons to be well measured. Recently
hadronic decays of charm mesons started being used to
study properties of scalar mesons, abundant products of these decays.

Charm decays have unique features, making them a very interesting tool for 
light quark spectroscopy: large couplings to scalar mesons and very small 
(less than 10\%) non-resonant components; an initial state which is always 
well defined: the spin-0 D meson; and a spectrum that is not constrained by 
isospin and parity conservation.

There are, however, 
some conceptual issues related to the formalism commonly used in the analysis 
of resonant substructure of hadronic decays: the correct representation
of overlapping broad states, which is closely connected to the issue of
formulating the unitarity constraint in three and 
four-body problems. Moreover, there is the question of 
how to relate the observations from charm to those from scattering.

In what follows I will briefly state the problem of the scalar mesons. Then I
will discuss how  we can use charm decays for new insights on the scalars. 
Finally, I will discuss the picture so far offered by hadronic decays of charm.

\section{The puzzling light scalars}

The light scalars are, in some sense, victims of their own simplicity.
Due to their broad widths and the lack of a distinctive angular distribution, 
the distinction
between scalar mesons and the non-resonant background is rather difficult.
Moreover, there are many overlapping states within a limited range of the mass
spectrum (up to 1.8 GeV). An additional difficulty is the fact that non-$q \bar
q$ states, like the lightest scalar glueball or multiquark states, all sharing
the same quantum numbers ($J^P=0^+$), are expected to populate the same region 
of the spectrum. We can say that the identification of the scalar mesons will
always be a difficult subject. Comprehensive reviews on scalar mesons can be
found in \cite{scalars} and references therein.

The main candidates, according to their isospin, are: 
 f$_0$(600) or $\sigma$(500), f$_0$(980), f$_0$(1370),
f$_0$(1500) and f$_0$(1710) ($I=0$);
$\kappa$(800) and $K^*_0$(1430) ($I=1/2)$;
a$_0$(980) and  a$_0$(1450) ($I=1$).

If all these states are confirmed, we have 19 states! Too many candidates to
fit even in two nonets.
While the actual existence of some of these states - the $\sigma$(500) and 
$\kappa$(800) - is controversial, other states just have poorly known 
parameters - f$_0$(980), a$_0$(980), f$_0$(1370). 
The interpretation of most of the scalar candidates is also controversial. 
Are they genuine $q \bar q$ mesons or more complex objects? Take the case of
the a$_0$(980), for instance. Its expected width is 500 MeV, according to the 
CQM, whereas the measured width is in the range 50-100 MeV. This fact
leads to the interpretation of this state as a $qq\bar q\bar q$.

The most problematic states are the isoscalars. In addition to the controverse
about the $\sigma$(500), the nature of the f$_0$(980), there is the issue of
the f$_0$ family above 1 GeV and a possible mixing with the scalar glueball
\cite{mix}. The remaining of this 
note is devoted to isoscalars and to what can we learn about them from charm 
decays. The $I=1/2$ states states including the $\kappa$(800),
will be addressed in the talk by Carla G\"obel, to
appear in these proceedings.

\section{Charm decays and light scalars}

Hadronic decays of charm mesons are a natural place to look for scalars, with
unique features that provide new and complementary insights on this problem.

Scalars are copiously produced in charm decays. In 3 and 4-body hadronic decays
of D mesons, one always has a $\pi \pi$, a $K\pi$ or a 
$KK$ pair, important decay modes of scalar mesons.
The quantum interference between broad scalars and the usually large
non-resonant background, which is a plague in scattering experiments, does not
affect charm decays because the non-resonant component is always very small.

But the most appealing features of D decays, when compared to scattering 
experiments, are related to the difference 
in the constraints that build the $\pi \pi$, $K\pi$ and $KK$ spectra.
In scattering
experiments, only the strong interaction is involved. The observed spectrum is
determined by the conservation of isospin and parity. Parity and isospin are
violated in D meson decays, where the observed spectrum is determined
by the quark content of the initial state, after the weak decay of the c quark.
 
It is illustrative to compare, for instance, in the $\pi^+ \pi^- \pi^+$ 
final state, the Dalitz 
plots from $D^+$, $D^+_s$ decays\cite{e791a,e791b} (see fig\ref{3pi})
and from $p \bar n$ annihilation\cite{obelix}, which is 
most similar to D decays. The differences due to production dynamics are
apparent at a glance. Comparing the $D^+$ and $D^+_s$ 
Dalitz plots we see clearly the effects of the different quark content of the 
initial state.
 
There is a related aspect which is also crucial: the bulk of the hadronic decay
widths can be explained by a model in which resonances couple directly to the D
meson. There is no need to add couplings to other states, like glueballs:
$q \bar q$ states alone seem to be enough to account for the observed rates. 
Take the decay $D^+_s \to K^-K^+\pi^+$ as a typical case. The main 
amplitudes are the external and internal W-radiation (see fig\ref{kkpi}). 
The decay modes 
corresponding to these amplitudes
are $D^+_s \to \phi \pi^+$ and $D^+_s \to K^* K^-$. These modes account for
almost  100\% of the $D^+_s \to K^-K^+\pi^+$ decay rate. The same argument
could be made using many other final states. Hadronic D decays are an extremely
complex process, and these types of quark diagrams are only an 
aproximation. This descripton, however, seems to work fairly well.
 
It is generally accepted that a resonance, being a real 
particle, must have the same parameters in whichever process it 
appears. The question one may ask is whether the states produced in
different processes are really the same. Consider, for instance, the
$f_0$(1370), $f_0$(1500) and $f_0$(1710) {\it imbroglio}. All three states 
have been observed by many experiments, with fairly well measaured parameters
(except for the $f_0$(1370)).
But according to the CQM, only two $q \bar q$ states are expected: one being 
mostly $s\bar s$ and another being mostly $u \bar u + d \bar d$. 
So, the three $f_0$'s could not belong to the same $q \bar q$ multiplet. 
 
Glueballs are expected to be produced in ''gluon-rich'' reactions, 
like central production, in addition to the genuine $q \bar q$ mesons.
Mixing between the scalar glueball and  the $q \bar q$ states is 
expected\cite{mix}. If this is really the case, then the observed states 
would be mixtures of $q \bar q$ and $gg$, rather than pure states.

On the other hand, in a ''gluon-poor'' reaction, like D decays,
glueballs are not expected to be produced. In D decays 
one would access directly the $q \bar q$ states with no mixing. 
In this case masses and widths measured in charm decays would be different
than those obtained in central production.
Also, the number of states present in D decays would be smaller.

One last aspect deserves some attention: the role of final state
interactions in charm decays. The Dalitz plots of charm decays 
can only be describe by models 
allowing interference between amplitudes
in which the resonance and the bachelor 
pseudo-scalar are in different states of relative orbital angular momentum.
The role the bachelor pseudo-scalar plays is decisive, which seems not to be
the case in $N\bar N$ annihilations.
In this sense Dalitz plot and partial wave analysis are not quite the same.
The case of the $D^+ \to K^- \pi^+ \pi^+$\cite{carlag} is typical.
We see in the Dalitz plot that the upper lobe of the 
$K^*(892)$ band is shifted  with respect to the lower one.
This effect is caused by the interference between the $l$=1  $D^+ \to K^*(892)
\pi^+$ and the $l$=0 amplitudes, like $D^+ \to K^*_0(1430) \pi^+$. 

We conclude this section by noting that relating results from scattering and 
charm decays is not so simple. The $D \to \pi\pi\pi$, for instance, cannot 
be explained on the basis of pure elastic $\pi\pi$ scattering. 
The energy dependent s-wave phase from $D \to \pi\pi\pi$ 
(or $K\pi\pi$) may not be the same as the $\pi\pi$ (or $K\pi$) phase 
shifts from peripheral hadron-hadron reactions.

\section{What have we learned so far from charm decays?}

There are only a few experimental results on light scalars from charm decays.
I will concentrate on the isoscalars: the $\sigma$ and the $f_0$ family.

\subsection{$\sigma$(500) or $f_0$(600)}

This is certainly the most controversial state. In charm decays it appears as 
an excess of signal events at low $\pi^+\pi^-$ mass. This effect is observed 
in the Dalitz plots of $D^+ \to \pi^+\pi^-\pi^+$, from E791\cite{e791b} and 
FOCUS, and of $D^0 \to \bar K^0 \pi^+\pi^-$, from CLEO\cite{cleo}.
The same structure was also observed in $J/\psi \to
\omega \pi^+\pi^-$ decay, from BES\cite{bes}. No such effect is observed in
$\pi^+ \pi^-$ scattering, where the $\sigma$ is interpreted 
not as a real particle, but as a dynamical threshold effect. 

The best description of charm decay data requires
the presence of a broad, scalar (in E791 analysis different spin assignments 
were also tested), complex amplitude at low $\pi^+\pi^-$
mass. A crucial aspect is that good fits can only be obtained allowing
the phase of this complex amplitude to vary across
the Dalitz plot. The above experiments have fitted their data assuming a 
Breit-Wigner function for this state, although it is known that for states like 
the $\sigma$ a Breit-Wigner is only an approximation. Different functional
forms may yield different values of mass and width. The very concept of mass 
and width is model dependent in this case. The CLEO Collaboration\cite{cleo} 
do not claim
evidence for the $\sigma$ meson due to the uncertainty in the best 
parameterization of this amplitude.
Anyway, good fits were obtained in all cases, and the values for the mass and 
width (see table \ref{sigma}) are
in good agreement - $M\sim$ 480 MeV, $\Gamma_0 \sim$ 320 MeV.

It would be interesting, definitely, to show the phase variation 
across the Dalitz plot without assuming any functional form for
the $\sigma$ amplitude. This is, unfortunately, very difficult because it 
involves a very large number of free parameters. 
In any case, it remains to be explained why in charm decays the $\sigma$
seems to be a real particle, but not in
low energy elastic $\pi^+\pi^-$ scattering.

\subsection{$f_0$(980)}

The width of this state is poorly known. The reason is that the $f_0$(980) seems
to behave differently depending on the reaction in which it is produced. While in
scattering it looks broader and with a large coupling to the $K \bar K$ channel,
in charm decays it looks just like a narrow regular $q\bar q$ resonance decaying
mostly into pions.
In the decay $D^+_s \to \pi^+\pi^-\pi^+$ the $f_0$(980)$\pi^+$ component 
correspond to over 50\% of the decay rate. 

E791 used a coupled channel Breit-Wigner (the Flatt\'e formula) in its fit
\cite{e791a}. The coupling to  $K\bar K$ channel was found to be consistent 
with zero.
An equally good fit was obtained using a regular Breit-Wigner, yielding 
$\Gamma_0 =$ (44$\pm$3) MeV. This is in agreement with preliminary results from
FOCUS ($\Gamma_0 \sim$ 55 MeV, from $D^+_s \to \pi^+\pi^-\pi^+$) and 
BES ($\Gamma_0 \sim$ 45 MeV, from $J/\psi \to \phi \pi^+\pi^-$).

The large rate in $D^+_s \to \pi^+\pi^-\pi^+$ suggests a strong affinity of the
$f_0$(980) with $s\bar s$, if we take the W-radiation amplitude to be the
dominant decay mechanism. In spite of a large $s\bar s$ in its wave function, 
the lack of a significant coupling to $K\bar K$ is due essentially to 
the narrow  $f_0$(980) width. 

The above situation reinforces the interpretation of this state as a 4-quark
state surrounded by a $K \bar K$ molecular cloud. At short distances, as in
D decays, we would access the $qq\bar q \bar q$ component, whereas in 
peripheral processes the molecular component would manifest itself.

\subsection{$f_0$(1370)/$f_0$(1500)}

The situation here is still rather confusing. The third state of the $f_0$
family above 1 GeV, $f_0$(1710), which would be mostly $s \bar s$, is 
difficult to access, since it lies near the edge of the $D^+_s$ decay phase 
space.

Charm decays are useful not only to measure the $f_0$(1370) and $f_0$(1500)
masses and widths, but also to infer the quark content of these 
two states. If both are $q \bar q$ resonances, both should appear in charm
decays. If, in addition, there is a significant $s \bar s$ component in their
wave function, these states should appear in the $D^+_s \to K^+K^-\pi^+$ decay.

Both E791 and FOCUS/E687, when analysing the $D_s^+ \to \pi^+\pi^-\pi^+$ 
decay, have found that only one state is necessary to describe the Dalitz 
plot, although they
do not agree on the measured parameters for this state. While E687 \cite{e687}
found a state with mass near 1475 MeV and a width of about 100 MeV (very similar
to FOCUS preliminary numbers, and very close to the well measured $f_0$(1500)
parameters), E791 \cite{e791a} found a somewhat wider state with a lower mass:
$M_0 =$ (1434$\pm$20) MeV, $\Gamma_0 =$ (172$\pm$32) MeV.

BaBar \cite{babar} have found no evidence of neither one of the $f_0$ states 
in the $D^0 \to \bar K^0 K^+ K^-$ decay.
FOCUS (see L. Edera's talk in these proceedings) have found a small component 
of $f_0$(1370)$\pi^+$ in the $D^+_s \to K^+K^-\pi$ decay, but the sum of 
all decay fractions is over 160\%. This is due to a large destructive 
interference, which is likely to be unphysical.

A large $D \to \pi^+ \pi^- \pi^+$  (and also $D \to \pi^+ \pi^0 \pi^0$) sample
are necessary to disentangle the $f_0$(1370)/$f_0$(1500) contribution.
Apparently  only one of the two $f_0$ would be a $q \bar q$ state (mostly $n \bar
n$), reinforcing the interpretation of the other one as the ground-state scalar
glueball.

\section{Conclusions}

The picture offered so far by charm decays points to the existence of two
scalar meson nonets, one having states with mass below 1 GeV and the other
with masses above 1 GeV.

In the low mass states we have the large rates of the $\sigma \pi$ in 
$D^+ \to \pi^+ \pi^- \pi^+$ and of the $f_0(980)\pi$ in 
$D^+_s \to \pi^+ \pi^- \pi^+$ decay as an indication that both are $q \bar q$, 
or, perhaps $qq \bar q\bar q$ states. 
The evidence for the neutral $\kappa(800)$
would be endorsed if evidence for the charged  $\kappa$ is also found. In the
cases of both $\sigma$ and $\kappa$, a demonstration of the phase variation 
would be very welcome.
It is also important to measure the $a_0$(980) width in either 
$D_s \to K^+K^-\pi$ or $D_s \to \bar K^0 K^+K^-$ decays.

In the region above 1 GeV more data is necessary to show which of the 
$f_0$'s are genuine $q \bar q$ states. Perhaps the answer is none of those 
observed in scattering experiments, since the mixing between 
the bare $q \bar q$ resonances and  the scalar glueball would not occur in charm
decays. Apparently only one state appears in charm decays, although it is not
clear yet which state this is. In any case, this state has no significant 
coupling to $K \bar K$.  

There are important conceptual issues to be addressed. The most important is to
formulate the unitarity constraint in multi-body decays. The
assumption of two-body elastic scattering as the basic process 
is not trivial and may not be justified. Even in the case of two-body 
elastic scattering, the introduction of a relative phase can
restore unitarity, which would be violated in models in which the amplitude is 
written as a sum of Breit-Wigners\cite{svec}.

Decays of charm mesons, with their unique features, offer a new way to look at 
the light scalar mesons.

\begin{table}[ht]
\centering
\caption{ \it Mass and width of the $\sigma$ assuming a Breit-Wigner model.}
\vskip 0.1 in
\begin{tabular}{|l|c|c|} \hline
 experiment  &  mass (MeV) & width (MeV) \\
\hline
\hline
 E791   & 478 $\pm$ 29 & 324 $\pm$ 46 \\
 CLEO   & 390 $\pm$ 60 & 282 $\pm$ 77 \\
 BES    & 513 $\pm$ 32 & 335 $\pm$ 67 \\
\hline
\end{tabular}
\label{sigma}
\end{table}

\vspace{1cm}

{\bf Figure 1 - {\it Dalitz plots from $D^+_s \to \pi^+ \pi^- \pi^+$ (left) and
      $D^+ \to \pi^+ \pi^- \pi^+$ (right) decays from Fermilab E791.}

\begin{figure}[hbt]

\includegraphics{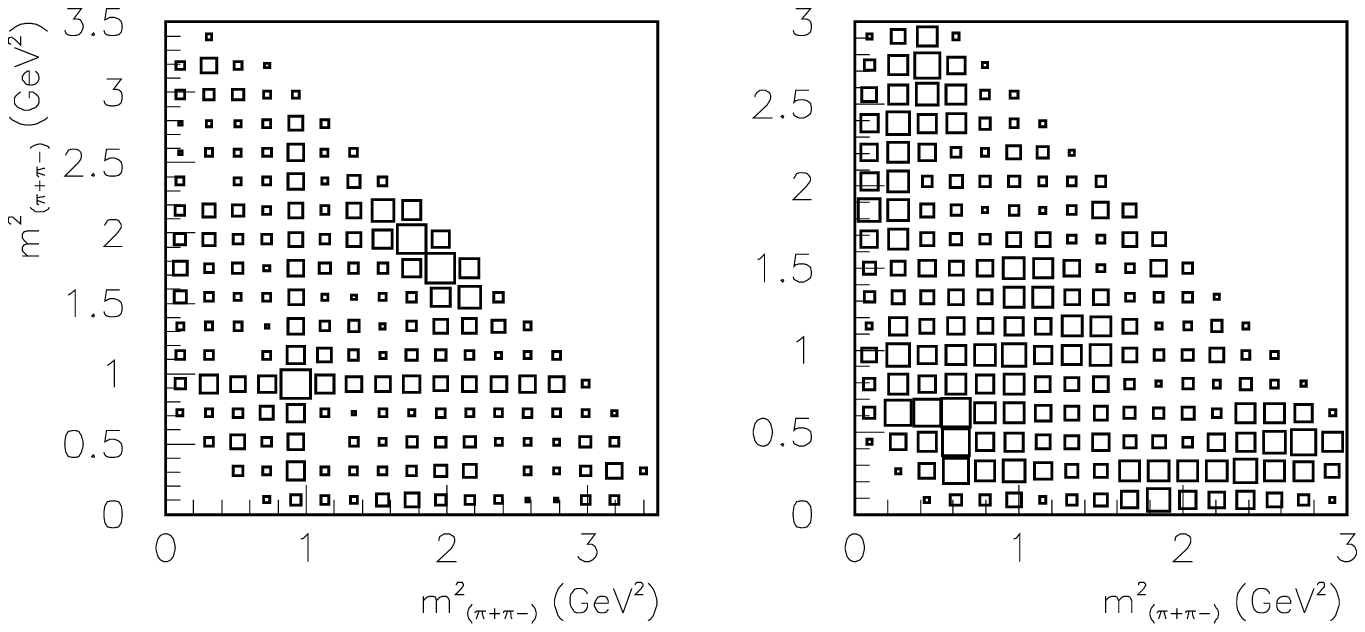}

   \label{3pi} 
\end{figure}

\vspace{7cm}      

{\bf Figure 2 - {\it Dominant amplitudes for $D^+_s \to K^-K^+\pi^+$ decay.}
\begin{figure}[hbt]
\includegraphics{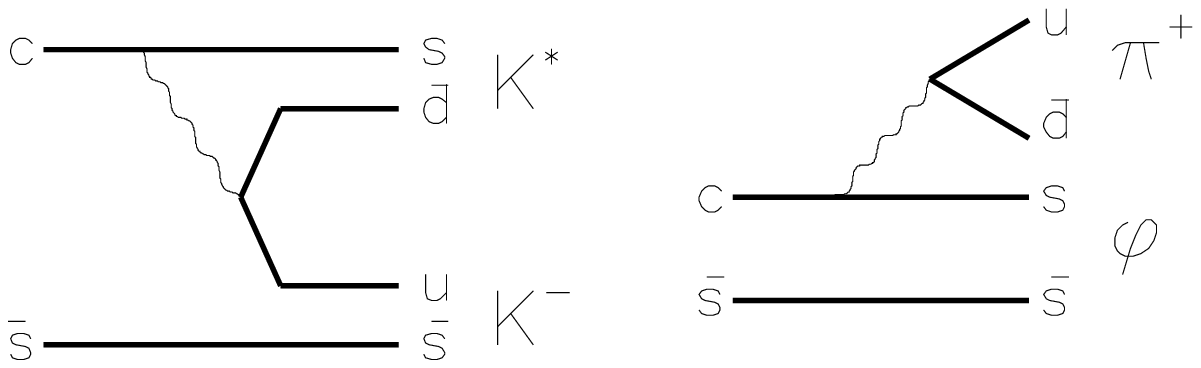}

    \label{kkpi} 
\end{figure}

\end{document}